\documentclass{article}
\usepackage{amssymb}
\input epsf.sty
\newenvironment{figMacPc}[4]				
{\begin{figure}[h]						% #1:nomefile della figura
	\epsfxsize=#2\centerline{\epsfbox{#1}}		% #2:dimen. orizz. figura
	\caption{#3}						% #3:didascalia
	\label{#4}							% #4:label
}
{\end{figure}}
\begin{document}
\title{A STOCHASTIC MODEL FOR THE STEPWISE MOTION IN ACTOMYOSIN DYNAMICS}
\author{\sc A. Buonocore \\ \sc A. Di Crescenzo \\ 
\sc B. Martinucci \\ \sc L.M. Ricciardi}
\date{\today}
%\subjclass{92C17, 60J70}
%
% 92C17: cell movement
% 60J70: applications of diffusion theory
%
%\keywords{Brownian motion, jumps, acto-myosin dynamics.}
\maketitle
\begin{abstract}
A jump-diffusion process is proposed to describe the displacements performed by single myosin heads along
actin filaments during the rising phases. The process consists of the superposition of a Wiener and a 
jump process, with jumps originated by sequences of Poisson-distributed energy-supplying pulses.
In a previous paper \cite{DCMR2002}, the amplitude of the jumps was described by a mixture of 
two Gaussian distributions. To embody the effects of ATP hydrolysis \cite{BR2002}, we now refine 
such a model by assuming that the jumps' amplitude is described by a mixture of three Gaussian distributions.
This model has been inspired by the experimental data of T.\ Yanagida and his co-workers concerning 
observations at single molecule processes level as described in \cite{KTIY1999} and 
in the references therein.


\end{abstract}
%----------------------------------------------------------------------------------
\section{Introduction}
\label{sec:1}
%----------------------------------------------------------------------------------
%
Muscle fibers are composed by a great number of even smaller fibers, called myofibrils, arranged 
parallel to the muscle fiber's major axis. The myofibrils consist of two different filaments: thick 
filaments, composed by myosin molecules, and thin filaments, composed by actin molecules.
Myosin molecules' extremity stick out from the filament as small heads.
Muscle contraction occurs when myosin molecules slide along actin filaments, fuelled by the chemical
energy originating from ATP hydrolysis. The traditional model explaining the mechanism of myosin movement 
is the so-called ``lever-arm swinging model'' in which the neck region of the myosin head swings to generate 
displacement (see Cooke \cite{Cooke}). 
The swing motion is coupled tightly to the ATP hydrolysis cycle i.e. the myosin molecule 
moves in a single forward step during each ATPase reaction. The size of myosin displacement 
is about $6 \rm{nm}$. 
\par
Using new techniques for manipulating single actin filaments, Kitamura {\em et al.\/} \cite{KTIY1999}
have obtained highly precise and reliable measurements of the displacements performed by single myosin heads along
actin filaments during their rising phases. They have recorded myosin displacements of $10 \rm{nm}\div{30 \rm{nm}}$. 
These values indicate that, during each biochemical cycle of ATP hydrolysis, the myosin head may interact 
several times with an actin filament, undergoing multiple steps. 
The steps occur randomly in time, mainly in a forward direction, roughly not more than $10\%$ 
in the backward direction. The step-size is approximately $5.5 \rm{nm}$, that identifies with the interval 
between adjacent monomers in one strand of the actin filament.
These observations clearly contradict the traditional lever-arm model, suggesting that a biased Brownian 
ratchet mechanism could be invoked: myosin head moves along the actin filament driven by 
the Brownian motion and the ATP hydrolysis biases the direction of the movement.
\par
In this paper we propose a stochastic model for the description of the displacements performed by
the myosin head durig a rising phase, on account of the experimental results presented in 
Kitamura and Yanagida \cite{KITA2001}. Our model is not exactly a ratchet-based model but 
all the relevant basic ingredients are preserved: exploitation of Brownian motion, 
the assumption that energy released by ATP hydrolysis is responsible for the myosin displacements 
and the intrinsic asymmetry of the actomyosin system.
To achieve directional motion, in addition to thermal noise and anisotropy, we consider an energy supply,
resulting from the hydrolysis of ATP (see Buonocore and Ricciardi \cite{BR2002}). 
We assume that this energy is stored by the myosin head and a part of it 
is released by small quanta of constant magnitudes. 
Usually not each energy release produces a step, but several releases are needed to generate a single step.
We further assume that myosin head is embedded in a viscous fluid in thermal equilibrium and denote by 
$\beta_v$ the viscous friction drag coefficient that in the experimental conditions is estimated as $90 \, \rm{pN}$.
\par
Aiming to  a phenomenological model to account 
for the available data on the displacements of the myosin head along an actin filament,
we take as starting point the articles by Di Crescenzo {\em et al.\/} \cite{DCMR2002}
and by Buonocore {\em et al.\/} \cite{BDCM:2002}, in which a stochastic process consisting of the superposition of 
Wiener and jump processes is discussed.
Myosin head's slide along the actin filament is viewed as a Brownian motion 
perturbed by jumps  that occurr according to a Poisson process.  
Their amplitude is described by a mixture of three Gaussian random variables. 
\par
The transition density of the stochastic process describing the stepwise motion
and some of its moments are obtained in Section \ref{sec:2}.
In Section~\ref{sec:3} we analyze the duration of the rising phase $U$ and the
position $V$ of the myosin at the end of the rising phase. For both variables we obtain the 
density and some moments.
In Section~\ref{sec:4} the special case when the backward jumps are not allowed is considered. 
Our results appear to 
be in qualitative agreement with the available experimental observations. 
%----------------------------------------------------------------------------------
\section{The model}
\label{sec:2}
%----------------------------------------------------------------------------------
%
Let $\{X(t),\,t\geq 0\}$ denote the real-valued stochastic process describing the motion of the
myosin head along the actin filament during the rising phase. 
In the absence of jumps, the myosin head is assumed to  move according to a Wiener process
with zero drift and infinitesimal variance $\delta^2 = 0.09 \, (\rm{nm})^2/\rm{ms}$. 
Here use of relation $\delta^2= 2 k_{B} T/\beta_{v}$ has been made, 
with $\beta_{v}=90 \, \rm{pN}$ and where $k_{B} T \sim 4\, \rm{pN nm}$ is the thermal 
energy at absolute temperature $T$, $k_{B}=0.01381\, \rm{pN}\cdot \rm{nm}/\rm{K}$ denoting 
the Boltzmann constant and $T=293\,\rm{K}$ the environmental temperature. 
According to the previous assumptions, at the occurence of the $i$-th event of a 
Poisson process $\{N(t),\,t\geq 0\}$ of intensity $\lambda$, 
an energy release occurs and consequently the myosin head performs a jump of random 
amplitude $Y_i$. The mean amplitude of such a jump is $L$, $0$ and $-L$ with 
probabilities $p$, $1-p-q$ and $q$, respectively. The following stochastic equation thus holds:  
\begin{equation}
  X(t)= \delta\, B(t)+\sum_{i=1}^{N(t)} Y_i,
  \qquad t>0,
\label{equation:1}
\end{equation}
where $\{B(t),\,t\geq 0\}$ is the standard Brownian motion, and $X(0)=0$. (We have arbitrarily set 
$0$ as the starting point of the motion). Note that the diffusive component of the motion has 
zero drift, according to the experimental evidence (see Kitamura {\em et al.\/} \cite{KTIY1999}). 
We suppose that processes $B(t)$, $N(t)$ and r.v.'s $Y_1,Y_2,\ldots$ are independent, 
where $Y_i$'s are copies of a random variable $Y$ characterized by distribution 
function $F_Y(y)$ and probability density $f_{Y}(y)$. The latter is assumed to be a mixture 
of three Gaussian densities $f_{Z_{1}}(y)$, $f_{Z_{2}}(y)$ and $f_{Z_{3}}(y)$ with 
means $L$, $0$ and $-L$ respectively, and identical variance $\sigma^2$. 
Hence, for all $y\in\mathbb{R}$ we have
\begin{eqnarray}
&& \hspace{0.15cm} f_{Y}(y)=\frac{{\rm d} F_{Y} (y) }{{\rm d}y}
=p \,f_{Z_{1}}(y)+(1-p-q) \,f_{Z_{2}}(y) 
+ q \,f_{Z_{3}}(y)
\nonumber \\
&& \hspace{1.15cm} = \frac{1}{\sqrt{2 \pi \sigma^{2} t}} \, \left\{p \exp\big(-\frac{(y-L)^{2}}{{2 \sigma^{2} t}}\big) 
+ (1-p-q) \exp\big(-\frac{y^{2}}{{2 \sigma^{2} t}}\big)\right.  
\nonumber \\
&& \hspace{1.15cm} \left. + q \exp\big(-\frac{(y+L)^{2}}{{2 \sigma^{2} t}}\big)  \right\},
\label{equation:6}
\end{eqnarray}
with ${\rm E}(Y)=L\,(p-q)$ and ${\rm Var}(Y)=\sigma^2$. 
We set $L=5.5\, \rm{nm}$ (the distance between adjacent actin monomers) and 
$\sigma^2=0.1\;(\rm{nm})^2/\rm{ms}$. 
Densities $f_{Z_{1}}(y)$, $f_{Z_{2}}(y)$ and $f_{Z_{3}}(y)$ account for the following: 
$f_{Z_{1}}(y)$ describes the size of forward steps, $f_{Z_{3}}(y)$ that of backward steps, 
whereas $f_{Z_{2}}(y)$ accounts for an energy release that does not produce a step. 
\par
For all $x\in\mathbb{R}$ and $t\geq{0}$ let us now introduce the following probability densities: 
\begin{eqnarray}
 && \hspace{-1.4cm} f(x,t) := {\partial \over \partial x}{\rm P}\{X(t)\leq x\,|\,X(0)=0\},
 \label{equation:2} \\
 && \hspace{-1.4cm} f_n(x,t) := {\partial \over \partial x}{\rm P}\{X(t)\leq x\,|\,X(0)=0, N(t)=n\},
\qquad n=1,2,\ldots\;.
\label{equation:3}
\end{eqnarray}
One has:
\begin{equation}
f(x,t)=\sum_{n=0}^{\infty}f_{n}(x,t)\,{\rm P}\{N(t)=n\},
\label{equation:4}
\end{equation} 
where 
\begin{equation}
{\rm P}\{N(t)=n\}
 =\frac{(\lambda t)^{n} \, 
{\rm e}^{- \lambda t}}{n!},
 \qquad n=0,1,\ldots,
\label{equation:9}
\end{equation}
is the probability distribution of a Poisson process  ${N(t)}$  having intensity $\lambda$. 
The conditional density $f_n(x,t)$ can be expressed as follows:
\begin{eqnarray}
&& \hspace{-2cm} f_n(x,t) = \left\{\begin{array}{ll}  
f_{W}(x,t\,|\,x_0) & \qquad \textrm{if $n=0$}
\vspace{0.5cm} \\
\displaystyle\int_{-\infty}^{\infty}f_{W}(x-y,t\,|\,x_0)\,{\rm d}F^{(n)}_{Y}(y) & \qquad \textrm{if $n=1,2,\ldots,$}
\end{array} \right.
\label{equation:5}
\end{eqnarray}
where 
$$ 
 f_{W}(x,t) 
 =\frac{1}{\sqrt{2\pi{\delta}^{2}t}} \,
 \exp\left\{-\frac{x^{2}}{2 \delta^{2} t}\right\}, 
 \qquad  x\in\mathbb{R}
$$
is the transition density of a Wiener process with zero drift and infinitesimal 
variance $\delta^2$, and where $F^{(n)}_{Y}(y)$ denotes the $n$-fold convolution of $F_{Y}(y)$ with itself. 
From (\ref{equation:6}), by induction we obtain:
\begin{equation}
{\rm d}F^{(n)}_{Y}(y)=\sum_{k=0}^{n} {n\choose k} (1-p-q)^{n-k}
\, \sum_{j=0}^{k} {k\choose j} \, 
p^{j} \, q^{k-j} \, \psi_{j,k,n}(y)\,{\rm d}y,
\label{equation:7}
\end{equation}
where $\psi_{j,k,n}(y)$ denotes a normal density with mean $(2j-k) L$ and variance $n \sigma^2$. 
Hence, from (\ref{equation:5}) and (\ref{equation:7}), for $x\in\mathbb{R}$, $t>0$ and $n=0,1,\ldots$ we have:
\begin{eqnarray}
&& f_n(x,t) =\frac{1}{\sqrt{2 \pi (\delta^{2}t+ n\,\sigma^2)}} 
\sum_{k=0}^{n} {n\choose k} (1-p-q)^{n-k} 
\nonumber \\
&& \hspace{2cm} \times \sum_{j=0}^{k} {k\choose j} p^{j} \, q^{k-j} 
\exp\left\{- \frac{[x-(2j-k)\, L]^2}{2(\delta^{2}t+ n\,\sigma^2)}\right\}. 
\label{equation:8} 
\end{eqnarray}
\par
Recalling Eq.\ (\ref{equation:4}), from Eqs.\ (\ref{equation:8}) and (\ref{equation:9}),
for all $x\in\mathbb{R}$ and $t\geq{0}$ we finally obtain the probability density of $X(t)$:
\begin{eqnarray}
&& \hspace{-0.8cm} f(x,t)= {\rm e}^{-\lambda t} \sum_{n=0}^{\infty} 
\frac{(\lambda t)^{n}}{n! \, \sqrt{2 \pi (\delta^2 t + n \,\sigma^2)}}
\, \sum_{k=0}^{n} {n\choose k} (1-p-q)^{n-k} 
\nonumber\\
&& \hspace{1.6cm} \times \sum_{j=0}^{k} {k\choose j} p^{j} q^{k-j} 
\, \exp\left\{- \frac{[x-(2j-k)\, L]^2}{2(\delta^{2}t+ n\,\sigma^2)}\right\}.
\label{equation:10} 
\end{eqnarray} 
Note that, as shown in Di Crescenzo {\em et al.\/} \cite{DCMR2002}, this is the solution of the 
following integro-differential equation: 
\begin{eqnarray*}
 {\partial\,f\over \partial t} \!\!\!\! 
 &=& \!\!\!\! -\lambda   \,f +{\delta^2\over 2}\,{\partial^2\,f\over \partial x^2} 
 + \lambda  \int_{-\infty}^{\infty}f(x-y,t)\,{\rm d}F_Y(y), 
\end{eqnarray*}
with initial condition $\displaystyle\lim_{t\downarrow 0} f(x,t)=\delta(x)$.
Density (\ref{equation:10}) is multimodal, with peaks located at multiples of $L$ (see the example plotted in Figure \ref{fig1}).
If $p>q$, at each point $x=kL$ ($k$ integer) where a peak exists, this peak is higher than 
the symmetric peak located at $-kL$, a manifestation of the prevalence of forward with respect
to backward displacements.
\begin{figMacPc} 
{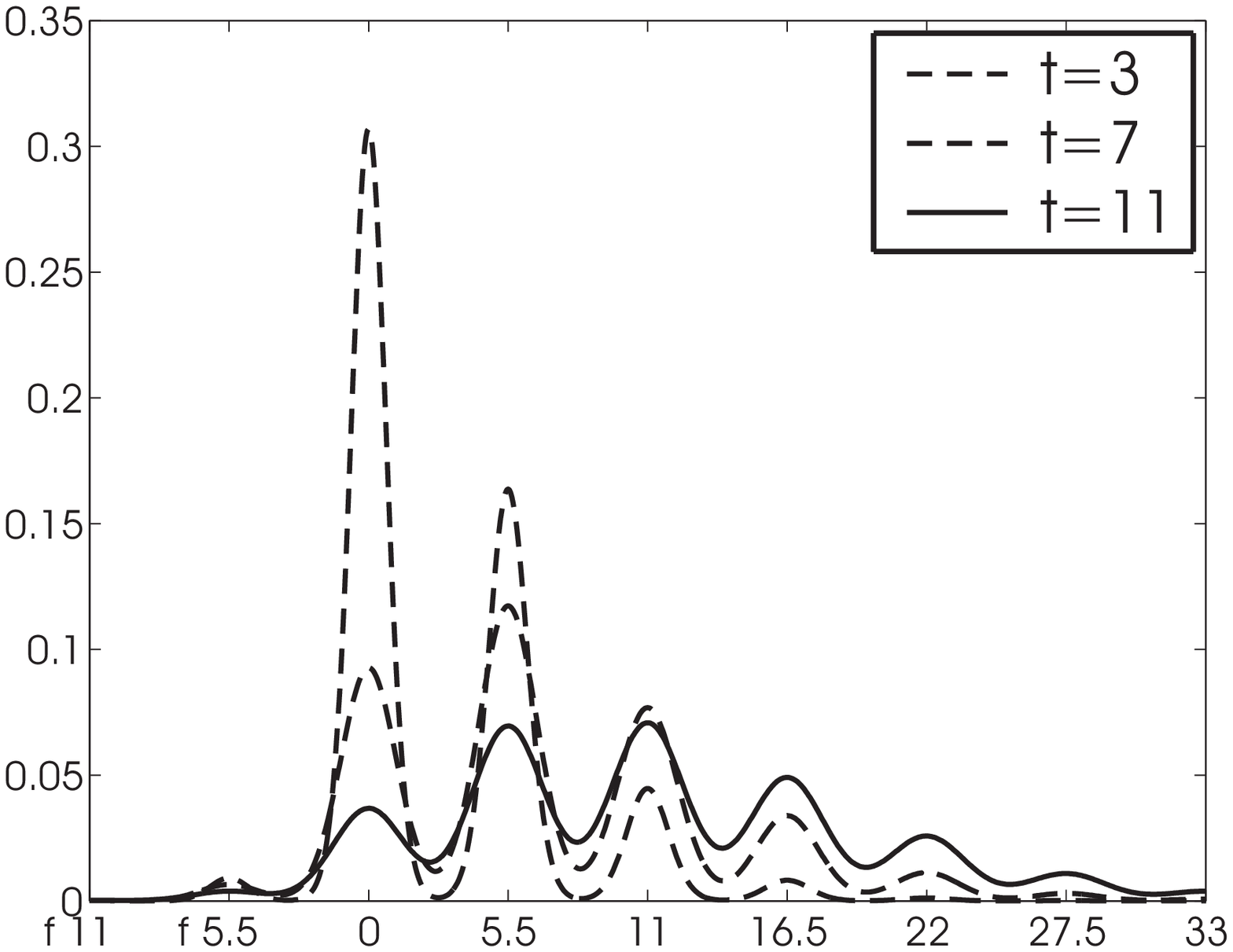}
{6cm}
{\small{Plot of density (\ref{equation:10}) with $\sigma^2=0.1$, $p=0.18$, $q=0.01$ and $\lambda=1.10$.}}
{fig1}
\end{figMacPc}
\par
Let us now obtain some moments of $X(t)$. From (\ref{equation:1}) we have: 
\begin{eqnarray}
&&\hspace{-0.8cm} {\rm E}[X(t)]:= m(t) = L(p-q) \, \lambda  t, 
\label{equation:11} \\
&&\hspace{-0.8cm} {\rm Var}[X(t)] := v^2(t) = 
\{[L^2 (p+q) +\sigma^2]\, \lambda   + \delta^2 \} \, t,
\label{equation:12} \\
&&\hspace{-0.8cm} {\rm E}\{[X(t)-m(t)]^3\} = L (p-q) (L^{2}+3\sigma^2)\, \lambda   t.
\label{equation:13}   
\end{eqnarray}
Hence, from (\ref{equation:11}) and (\ref{equation:12})  
the coefficient of variation follows: 
\begin{equation}
{\rm CV}[X(t)]:= \frac{v(t)}{m(t)}
=\frac{\sqrt{[L^2 (p+q) +\sigma^2]\, \lambda  + \delta^2} \,\, \sqrt{t}}
{{L(p-q) \lambda} \, t}.
\label{equation:14}
\end{equation}
Let us denote by $\mu(t)=X(t)/t$ the velocity of the myosin head displacement. 
Making use of Eqs.\ (\ref{equation:11}) and (\ref{equation:12}) we obtain the mean and the variance of $\mu(t)$:
\begin{eqnarray}
&& \hspace{-0.5cm} {\rm E}[\mu(t)] = L(p-q) \lambda,
 \label{equation:16} \\
&& \hspace{-0.5cm} {\rm Var}[\mu(t)] = \frac{{[L^2 (p+q) +\sigma^2]\, \lambda  + \delta^2 }}{t}.
 \nonumber
\end{eqnarray}
Finally, Eqs.\ (\ref{equation:12}) and (\ref{equation:13}) yield the skewness:
\begin{equation}
\frac{{\rm E}[X(t)-m(t)]^3}{v^3(t)}
=\frac{\lambda (p+q) L (p-q) (L^{2}+3\sigma^2)}
{\{[L^2 (p+q) +\sigma^2]\lambda + \delta^2 \}^{3/2} \,\, \sqrt{t}}.
\label{equation:15}
\end{equation}
The skewness goes to zero as $t\to+\infty$, it is positive (negative) as $p>q$ ($p<q$),
whereas it vanishes when $p=q$. 
Under the assumption $p>q$, which is of interest in the context of myosin head's motion,
we have the following monotonicity properties of the skewness:
\begin{itemize}
\item[(i)]  
as a function of $\lambda\in (0,+\infty)$, it is increasing for 
$\lambda<\overline\lambda\equiv 2\delta^2/ [L^2 (p+q)+\sigma^2]$,  
decreasing for $\lambda>\overline\lambda$, and goes to $0$ as 
$\lambda\to +\infty$;
\item[(ii)]  
as a function of $\sigma^2\in (0,+\infty)$, it is increasing if 
$\sigma^2<\overline\sigma^2\equiv L^2 [2\,(p+q)-1]+2\delta^2/\lambda$, 
decreasing for $\sigma^2>\overline\sigma^2$, and tends to $0$ when $\sigma^2\to +\infty$. 
\end{itemize}
\par
Let us denote by $M$ the r.v.\ describing the number of energy quanta releases required to produce 
one step, that corresponds to a jump of mean amplitude $-L$ or $L$.
On the ground of our assumptions, $M$ is a geometric r.v.\ with parameter $p+q$.
Denoting by $D$ the r.v.\ describing the dwell time, i.e. the time between consecutive myosin jumps, we have
\begin{equation}
D=T_{1}+\cdots+T_{M},
\label{equation:8bis}
\end{equation}
where $T_i$ is the duration of the effect of the $i$-th energy quantum release. 
Being $N(t)$ a Poisson process with intensity $\lambda$, we have that 
$T_1, T_2,\ldots$ are i.i.d.\ exponential r.v.'s 
with mean $\lambda^{-1}$. This is in agreement with Kitamura and Yanagida \cite{KITA2001}, whose dwell-time histograms   
are well fitted by exponential curves. 
From (\ref{equation:8bis}) we obtain  
$$
{\rm E}({\rm e}^{D s})=\sum_{k=1}^{+\infty} {\rm E} ({\rm e}^{s(T_1+\cdots+T_k)}) \, 
{\rm P} (M=k) = \frac{\lambda (p+q)}{\lambda (p+q) -s},  \qquad s<\lambda (p+q),
$$
so that the dwell time $D$ is exponentially distributed with mean 
\begin{equation}
{\rm E}(D)= \frac{1}{\lambda (p+q)}.
\label{equation:8tris}
\end{equation}
From (\ref{equation:16}) and (\ref{equation:8tris}) we have 
$$
 {\rm E}[\mu(t)] = {p-q\over p+q}\, {L\over {\rm E}(D)},
$$
so that the mean velocity ${\rm E}[\mu(t)]$ can be seen as the mean net displacement 
in one step divided by the mean time needed to have one step. 
%----------------------------------------------------------------------------
\section{Rising phase}
\label{sec:3}
%----------------------------------------------------------------------------
%
Some histograms of the number of steps per displacement have been shown in Kitamura and Yanagida \cite{KITA2001}
and Kitamura {\em et al.\/} \cite{KTIY1999}. There are $66$ observed displacements under a 
near-zero load and $77$ displacements for a load between $0$ and $0.5$\, $\rm{pN}$.
The number of net steps in each displacement ranges from 1 to 5 under a near-zero load 
and from 1 to 4 for a load between 0 and 0.5\, $\rm{pN}$. 
Moreover, the distribution of the total number of net steps observed in a rising phase 
minus 1 is well-fitted by a Poisson random variable.
Hence, in order to include such features in our model we introduce a random variable 
$U$ that describes the duration of the myosin head rising phase. 
Let $\{N_+(t),\,t\geq 0\}$ and $\{N_-(t),\,t\geq 0\}$ be two independent Poisson processes 
characterized by intensities $\lambda p$ and $\lambda q$, that describe the numbers of 
forward and backward steps performed by the myosin head during a rising phase, respectively. 
We assume that $U$ is the first-passage time of the number of net steps, given by 
$R(t):=N_+(t)-N_-(t)$, through a Poisson-distributed random threshold $S$, with  
\begin{equation}
{\rm P}(S=k) =\frac{{\rm e}^{-\rho} \, {\rho}^{k-1}}{(k-1)!}, \qquad k=1,2,\ldots,
\label{equation:15bis}
\end{equation}
where $\rho>0$ (note that this assumption refines a previous model considered in 
Buonocore {\em et al.\/} \cite{BDCM:2002}). The mean number of net steps during 
the rising phase is thus given by 
\begin{equation}
{\rm E}(S)=\rho+1.
\label{equation:16bis}
\end{equation}
We note that $R(t)$ is a randomized random walk, whose transition probability 
for positive $\lambda$, $p$ and $q$ is given by (see, for instance, Conolly \cite{Co75})
$$
 p_{k}(t)={\rm P}\{R(t)=k\,|\,R(0)=0\}
 ={\left(\frac{p}{q}\right)}^{k/2}e^{-\lambda(p+q)t}\,
 I_{k}(2\lambda t\sqrt{pq}),
 \qquad t>0,
$$
where $I_{k}(x)=\sum_{n=0}^{+\infty} {\left({x/2}\right)}^{2n+k}/{[n!(n+k)!]}$ is the modified Bessel function.
For all $t>0$, the p.d.f of $U$ can be expressed as follows:
\begin{equation}
f_{U}(t) = \sum_{k=1}^{+\infty} f_{U|S}(t\,|\,k) \, {\rm P}(S=k),
\label{equation:17}
\end{equation}
where ${\rm P}(S=k)$ is given in (\ref{equation:15bis}) and $f_{U|S}(u\,|\,k)$ is the 
first-passage-time density of $R(t)$ through $k$. The latter is given by  
\begin{equation}
 f_{U|S}(t\,|\,k)={k\over t}\,p_{k}(t)
 ={k\over t}\,{\left(\frac{p}{q}\right)}^{k/2}e^{-\lambda(p+q)t}\,
 I_{k}(2\lambda t\sqrt{pq}).
 \qquad t>0,
\label{equation:22}
\end{equation}
From (\ref{equation:15bis}), (\ref{equation:17}) and (\ref{equation:22}) we have 
\begin{equation}
 f_{U}(t) 
 = {e^{-\lambda(p+q)t}\over t}{\rm e}^{-\rho}\sum_{k=1}^{+\infty} 
 {k\over (k-1)!}{\left(\frac{p}{q}\right)}^{k/2}\,{\rho}^{k-1}I_{k}(2\lambda t\sqrt{pq}),
 \qquad t>0.
\label{equation:23}
\end{equation}
Making use of Eq.\ (\ref{equation:23}) and of Eqs.\ 4.16.1 and 4.16.2 of 
Erd\'elyi {\em et al.\/} \cite{EMOT54}, for $p>q$ we obtain:
\begin{eqnarray}
&& \hspace{-0.8cm} {\rm E}(U)={\rho+1\over \lambda \,(p-q)}, 
 \label{equation:26} \\
&& \hspace{-0.8cm} {\rm Var}(U)={(2 \rho+1)p+q\over \lambda^{2} \, (p-q)^{3}}, 
 \nonumber  \\
&& \hspace{-0.8cm} {\rm CV}(U)={1\over \rho+1}\,\sqrt{(2 \rho+1)p+q\over p-q}.
 \nonumber
\end{eqnarray}
\begin{figMacPc} 
	{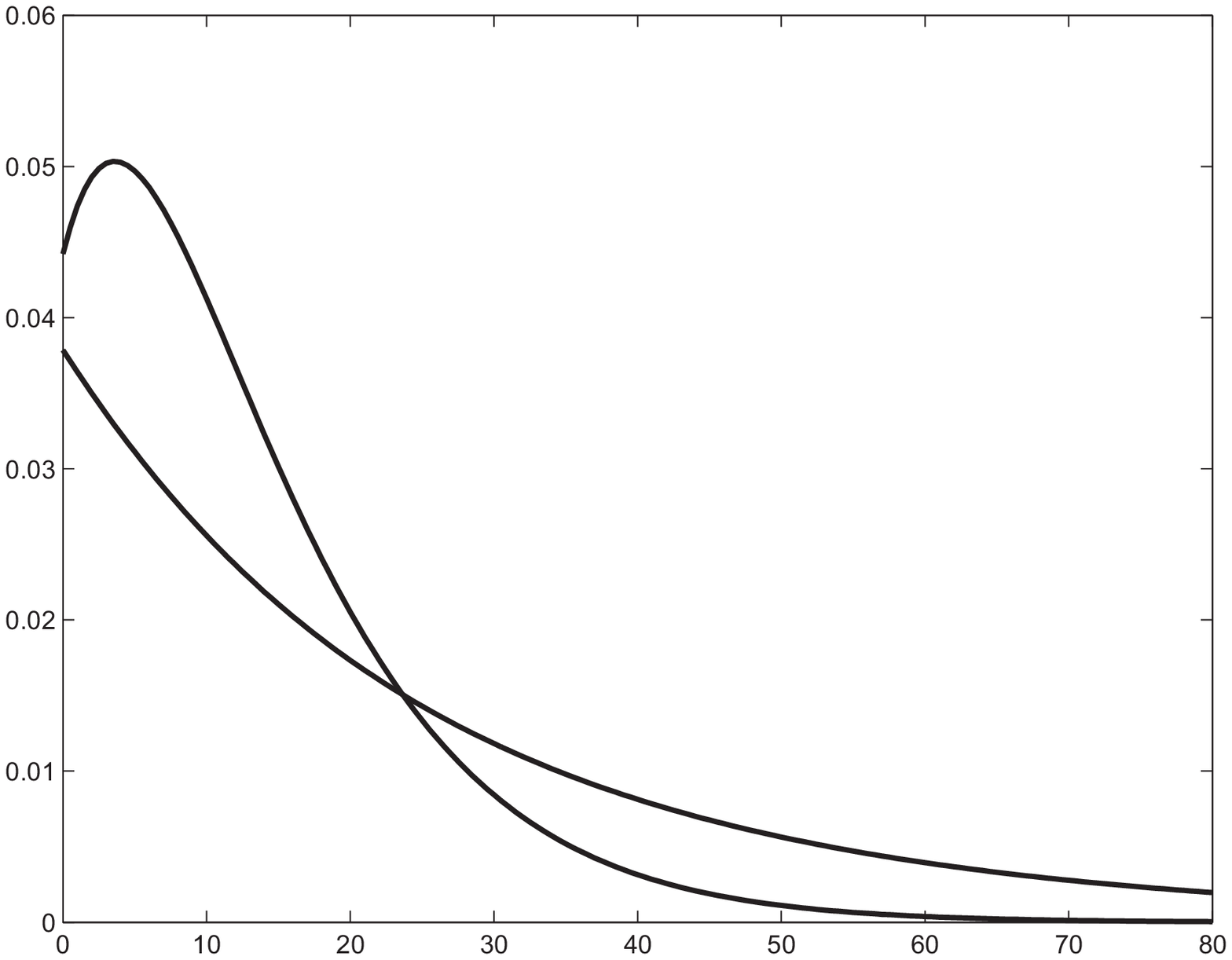}
	{6cm}
	{\small{Plots of density (\ref{equation:23}) in the following cases: 
		(i)  $\rho=1.5$, $[\lambda(p+q)]^{-1}=4.8\,$ms and $p/q=18$, where 
		${\rm E}(U)=13.37\,$ms and ${\rm Var}(U)=122.80\,$ms$^2$, and 
		(ii) $\rho=0.6$, $[\lambda(p+q)]^{-1}=12.2\,$ms and $p/q=6$, where 
		${\rm E}(U)=27.83\,$ms and ${\rm Var}(U)=858.98\,$ms$^2$.}}
	{fig5}
\end{figMacPc}
\newline
Note that ${\rm E}(U)={\rm E}(S)/{\rm E}[R(t)/t]$, so that the mean duration of the 
rising phase equals the mean number of net steps times the net steps rate.  
Two plots of density (\ref{equation:23}) are shown in Figure 2. 
Parameters $p$ and $q$ are chosen according to the experimental results 
concerning myosin number of steps along the actin filament for different 
load values that are given in Kitamura and Yanagida \cite{KITA2001}, where the ratio between the number of forward steps 
and backward steps is $18$ under a near-zero load, and is $6$ for a load between $0$ and $0.5$ pN. 
Moreover, recalling from (\ref{equation:8tris}) that the mean dwell-time is $[\lambda (p+q)]^{-1}$,
$\lambda$ is obtained from 
\begin{eqnarray*}
&& \hspace{-2cm} [\lambda (p+q)]^{-1} = \left\{\begin{array}{ll}  
4.8 \,\rm{ms} & \qquad \textrm{under a near-zero load} \\
12.2 \,\rm{ms} & \qquad \textrm{for a load between $0$ and $0.5$ pN}
\end{array} \right.
\end{eqnarray*}
where the rates of dwell times, appearing on the right-hand sides, are experimentally evaluated 
in the presence of 1 $\rm{\mu M}$ ATP and at $20\,\rm{C}$ (see Kitamura {\em et al.\/} \cite{KTIY1999} 
and Kitamura and Yanagida \cite{KITA2001}). Moreover, due to (\ref{equation:16bis}), the value of $\rho$ 
is chosen according to the following experimental mean numbers of steps 
observed during rising phases (see Kitamura and Yanagida \cite{KITA2001}):
\begin{eqnarray*}
&& \hspace{-2cm} \rho+1 = \left\{\begin{array}{ll}  
2.5 & \qquad \textrm{under a near-zero load} \\
\displaystyle 1.6 & \qquad \textrm{with a load between $0$ and $0.5\, \rm{pN}$.}
\end{array} \right.
\end{eqnarray*} 
Hence, in case (i) of Figure 2, where $f_{U}(t)$ exhibits 
a positive mode, it is $p/q=18$, $\lambda (p+q)=(4.8)^{-1}$ and $\rho=1.5$; in case (ii), where 
$f_{U}(t)$ is strictly decreasing, we have $p/q=6$, $\lambda (p+q)=(12.2)^{-1}$ and $\rho=0.6$. 
\par
Let us now denote by $V$ the random variable denoting the  
position attained by the myosin head at the end of the 
rising phase. Its density is given by:
\begin{equation}
 f_{V}(x):= \int_{0}^{+\infty} f(x,t)\, f_{U}(t)\, {\rm d}t, 
 \qquad x\in\mathbb{R}
\label{equation:27}
\end{equation}
with $f(x,t)$ expressed in (\ref{equation:10}) and $f_{U}(t)$ given in (\ref{equation:23}). 
In order to provide a qualitative insight of its behaviour, in 
Figures $3$ and $4$ we show two instances of probability density (\ref{equation:27}).
\begin{figMacPc} 
	{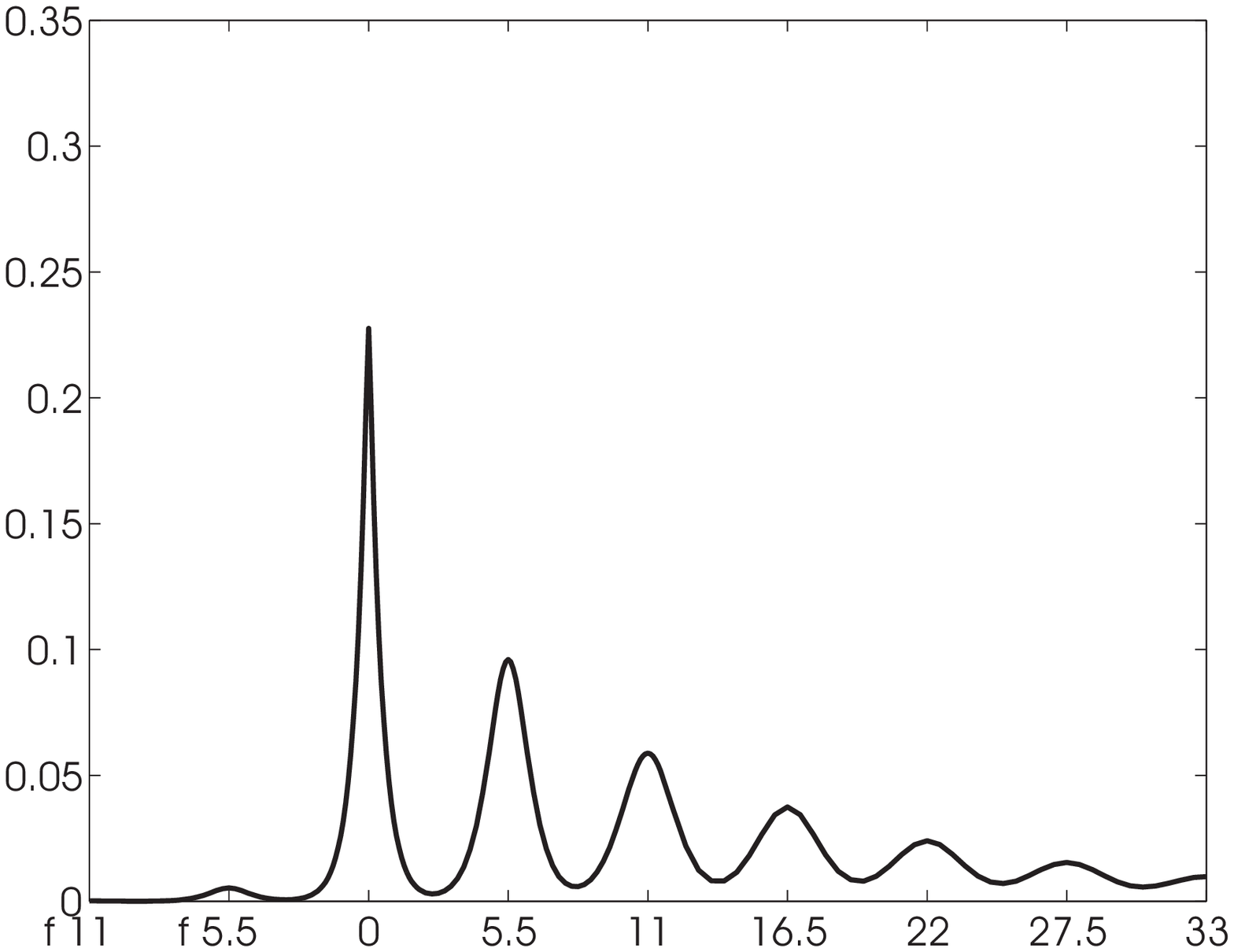}
	{6cm}
	{\small{Plot of density (\ref{equation:27}) with $\rho=1.5$, $\sigma^2=0.1$, $p=0.72$, 		$q=0.04$ and $\lambda=0.27$.}}
	{fig6}
\end{figMacPc}
\begin{figMacPc} 
	{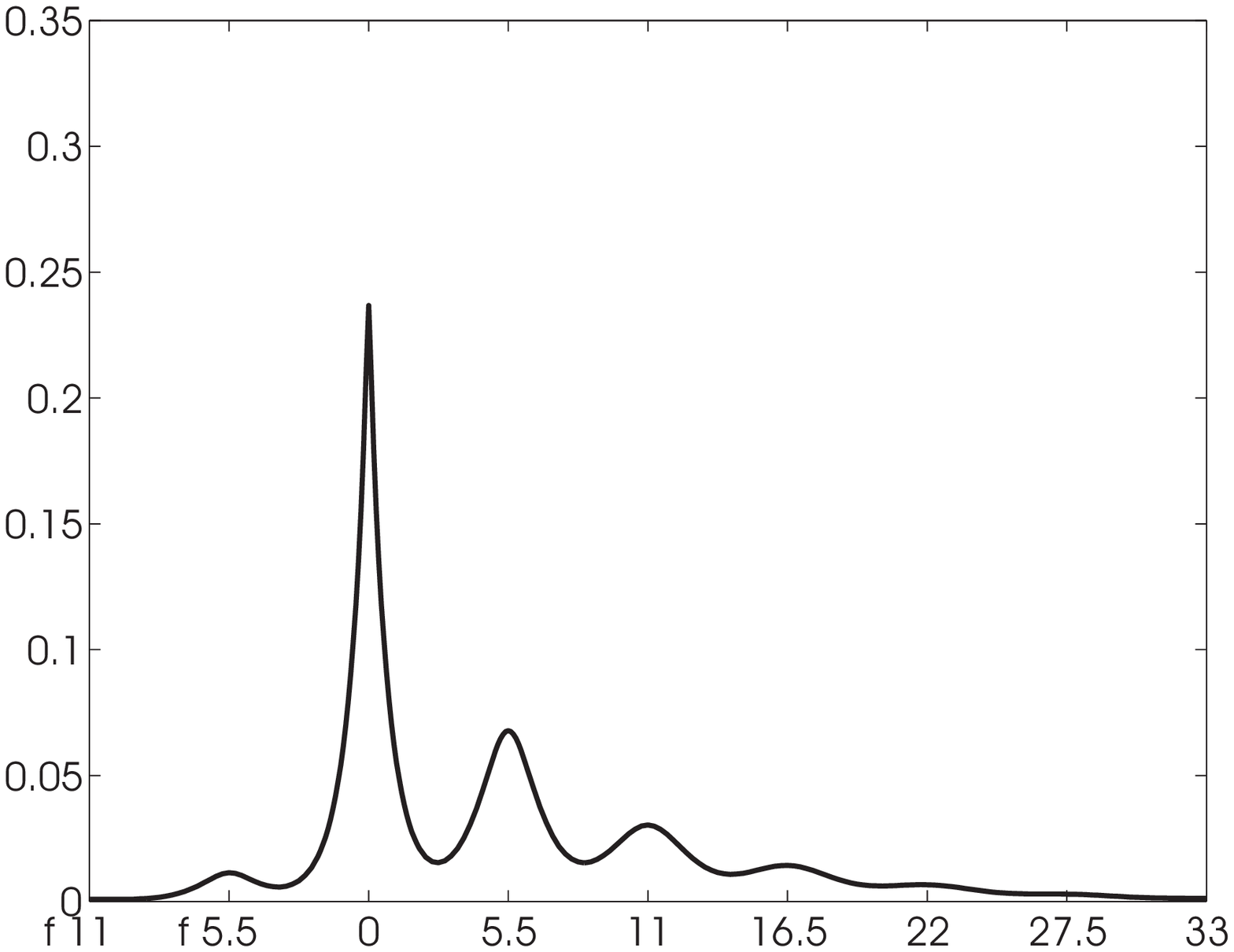}
	{6cm}
	{\small{Plot of density (\ref{equation:27}) with $\rho=0.6$, $\sigma^2=0.1$, 
		$p=0.18$, $q=0.03$, $\lambda=0.39$.}}
	{fig7}
\end{figMacPc}
For $p>q$, making use of Eq.\ (\ref{equation:27}), we have:
\begin{eqnarray}
&& \hspace{-0.2cm} {\rm E}(V) = (\rho+1) \, L, 
 \label{equation:24} \\
&& \hspace{-0.2cm} {\rm Var}(V) = \frac{\lambda L^2\,[(2\rho+1)\,p+q] + (\rho+1) \{\lambda[L^2 (p+q) + \sigma^2] + \delta^2\}}{\lambda (p-q)},
 \label{equation:25} \\
&& \hspace{-0.2cm} {\rm CV}(V) =\frac{\sqrt{\lambda L^2\,[(2\rho+1)\,p+q] + (\rho+1) \{\lambda[L^2 (p+q) + \sigma^2] + \delta^2\}}}
{(\rho+1) L \sqrt{\lambda (p-q)}}.
 \nonumber
\end{eqnarray}
Comparing Eq.\ (\ref{equation:24}) with (\ref{equation:26}) and (\ref{equation:16}) we have 
${\rm E}(V)={\rm E}(U)\cdot {\rm E}[\mu(t)]$, so that the mean position at the end of the 
rising phase equals the mean duration of the rising phase times the constant mean velocity 
of the myosin head. In Tables \ref{tab:1} and \ref{tab:2} the values of ${\rm E}(V)\pm 2\sqrt{{\rm Var}(V)}$
are indicated. These are obtained via Eqs.\ (\ref{equation:24}) and (\ref{equation:25})
for some choices of $p$, $q$ and $\lambda$ under the conditions $p/q=18$, $\rho=1.5$ and $E(V)=13.75$
(Table \ref{tab:1}) and $p/q=6$, $\rho=0.6$ and $E(V)=8.8$ (Table \ref{tab:2}).
\begin{table}[tc]
\centering
\caption{See text for explanation}
\vfill
\label{tab:1}  
\begin{tabular}{ccccc}
\hline\noalign{\smallskip}
$p$ & $q$ & $\lambda$ & ${\rm E}(V)-2\sqrt{{\rm Var}(V)}$ & ${\rm E}(V)+2\sqrt{{\rm Var}(V)}$\\
\noalign{\smallskip}\hline\noalign{\smallskip}
0.09 &0.005 & 2.19 & -15.82 & 43.32 \\
 0.18 & 0.01 & 1.10 & -15.72 & 43.22 \\
 0.36 & 0.02 & 0.55 & -15.67 & 43.17 \\
 0.72 & 0.04 & 0.27 & -15.65 & 43.15 \\
\noalign{\smallskip}\hline
\end{tabular}
\end{table}
\begin{table}[hc]
\centering
\caption{See text for explanation}
\vfill
\label{tab:2}  
\begin{tabular}{ccccc}
\hline\noalign{\smallskip}
$p$ & $q$ & $\lambda$ & ${\rm E}(V)-2\sqrt{{\rm Var}(V)}$ & ${\rm E}(V)+2\sqrt{{\rm Var}(V)}$\\
\noalign{\smallskip}\hline\noalign{\smallskip}
0.03 & 0.005 & 2.34 & -16.70 & 34.30 \\
0.06 & 0.01 &  1.17 & -16.45 & 34.05\\
0.12 & 0.02 & 0.59 & -16.32 & 33.92\\
0.24 & 0.04 & 0.29 &  -16.26 & 33.86\\
\noalign{\smallskip}\hline
\end{tabular}
\end{table}
%
%----------------------------------------------------------------------------
\section{Rising phase in the absence of backward steps}
\label{sec:4}
%----------------------------------------------------------------------------
%
The experimental results of Kitamura and Yanagida \cite{KITA2001} and Kitamura {\em et al.\/} \cite{KTIY1999} 
show that backward steps during a rising phase are rare with respect to forward steps. 
Hence, it is challenging to analyse the results given in the previous section when backward 
steps are not allowed, i.e.\ by assuming $q=0$. In this section we thus assume that $R(t)$ identifies 
with $N_+(t)$, which is a Poisson process with intensity $\lambda\,p$. Hence, the random 
duration $U$ of the rising phase in the absence of backward steps is the first-passage 
time of $N_+(t)$ through the random threshold $S$. Recalling Eq.\ (\ref{equation:8tris}), 
in this case the dwell time is exponentially distributed with mean value 
$(\lambda p)^{-1}$, so that $f_{U|S}(t\,|\,k)$ is an Erlang density with parameters $k$ 
and $(\lambda p)^{-1}$:
\begin{equation}
f_{U|S}(t\,|\,k) = \lambda p \, {\rm e}^{-\lambda p t} \, 
\frac{(\lambda p t)^{k-1}}{(k-1)!}, \qquad t>0.
\label{equation:31}
\end{equation}
\par
From Eqs.\ (\ref{equation:15bis}), (\ref{equation:17}) and (\ref{equation:31}), we obtain
the probability density of the rising phase duration in the absence of backward steps:
\begin{equation}
f_{U}(t)={\rm e}^{-\rho} \, \lambda p\, {\rm e}^{-\lambda p t}
I_{0}\Big(2 \sqrt{\rho \, \lambda p \, t}\Big), \qquad t>0.
\label{equation:32}
\end{equation}
From (\ref{equation:32}) we have that $f_{U}(0)={\rm e}^{-\rho} \lambda p >0$, and that 
$f_{U}(t)$ is a Polya frequency of order 2 density (${\rm PF}_2$), i.e.\ a logconcave density 
(see Marshall and Olkin \cite{MaOl1979}). This means that $[U-t\,|\,U>t]\geq_{\rm lr}[U-\tau\,|\,U>\tau]$ 
whenever $0<t\leq \tau$ (see Theorem 1.C.22 of Shaked and Shantikumar \cite{ShSh94}). In other 
words, as time goes on, the residual time of rising phase decreases in the likelihood ratio order 
sense. Moreover, from (\ref{equation:32}) we obtain
$$
\hspace{0.3cm} \frac{{\rm d}\, f_{U}(t)}{{\rm d}t}={\rm e}^{-\rho} \, (\lambda p)^{2} \, {\rm e}^{-\lambda p t}
\left[ (\rho-1) I_{0}\Big(2 \sqrt{\rho \, \lambda p \, t}\Big) - \rho\, 
I_{2}\Big(2 \sqrt{\rho \, \lambda p \, t}\Big)\right], 
 \qquad t>0.
$$
From this expression it is possible to show that density $f_{U}(t)$ is 
decreasing  for all $t>0$ if $\rho\leq{1}$, whereas density (\ref{equation:32}) is unimodal 
with a positive mode if $\rho>1$. 
%
%--------------------------------------------------------------------
\subsection*{\bf Acknowledgments}
Work performed within a joint cooperation agreement between
Japan Science and Technology Corporation (JST) and Universit\`a di Napoli
Federico II, under partial support by INdAM (G.N.C.S.).
%--------------------------------------------------------------------
%
\bibliography{}

\noindent
{\em A.\ Buonocore, B.\ Martinucci and L.M.\ Ricciardi:} 
{\sc Dipartimento di Matematica e Applicazioni, Universit\`a di Napoli Federico II, 
Via Cintia, Napoli I-80126, Italy}\\
{\em A.\ Di Crescenzo:} 
{\sc Dipartimento di Matematica e Informatica, Universit\`a di Sa\-lerno, 
Via S.\ Allende, I-84081  Baronissi (SA), Italy}
\end{document}